# Repulsive van der Waals forces due to hydrogen exposure on bilayer Graphene

Mathias Boström[1,2] and Bo E. Sernelius[1,*]

[1]Division of Theory and Modeling, Department of Physics, Chemistry and Biology, Linköping University SE-581 83 Linköping, Sweden, [2]Department of Applied Mathematics, Australian National University, Canberra, Australia.

We consider the effect of atomic hydrogen exposure to a system of two undoped sheets of graphene grown near a silica surface (the first adsorbed to the surface and the second freestanding near the surface). In the absence of atomic hydrogen the van der Waals force between the sheets is attractive at all separations causing the sheets to come closer together. However, with addition of atomic hydrogen between the sheets the long range van der Waals interaction turns repulsive at a critical concentration. The underlying triple layer structure ($SiO_2$ –Atomic Hydrogen Gas –Air) gives rise to a long range repulsion that at large enough separations dominates over the more rapidly decaying attraction between the two-dimensional undoped graphene sheets (and between the outer graphene sheet and $SiO_2$). This may be an avenue to tune the separation between two graphene sheets with the gas concentration. Doping of the graphene layers increases the attractive part of the interaction and hence reduces the net repulsive interaction.

# 1. Introduction

The Casimir and Lifshitz theories of dispersion forces [1-8] have occupied such a vast literature that little should remain to be said. However, we will demonstrate a previously unrecognized, and for nano-technological applications potentially important effect, which is due to competition between attractive and repulsive van der Waals forces. Attractive and repulsive van der Waals forces between triple layer films in general are well known [5-9] (see also some discussion in the concluding section.) We consider a system with two undoped two-dimensional (2D) graphene sheets grown near a silica surface. The two sheets, the first adsorbed to the $SiO_2$ surface and the second freestanding some nanometers from the surface, will in its original configuration feel strong attractive van der Waals forces that would drive the bilayer system together and ultimately make both adsorb at the silica surface. However, if the system is exposed to atomic hydrogen gas between the sheets the interaction at a specified separation turns repulsive at a critical hydrogen concentration. Conversely, the interaction at fixed, high enough, concentration turns repulsive at a critical large enough separation. This means that the outer sheet in the presence of a critical concentration of atomic hydrogen floats freely several nanometers away from the surface. Doping of the graphene layers increases the attraction and reduces the region of repulsion.

Some of the highly interesting material properties of graphene were first discussed in the literature by Boehm et al. [9] in 1962. With modern technology it is now possible to produce large-area, both freestanding and adsorbed, graphene sheets, and graphene has become one of the most advanced 2D materials of today. Due to its superior transport properties and having a mobility that is insensitive to variations of carrier density and temperature it has a high potential for technological applications [10-15]. Bi-layer graphene has been exploited to study current drag between two two-dimensional sheets (see ref. [16] and similar work done for bilayer 2D quantum wells [17,18]). The current drag depends on the distance between the two 2D sheets. This distance can according to our modeling be tuned with addition of atomic hydrogen atoms, hydrogen molecules or helium atoms. Due to the importance of intercalated atomic hydrogen we focus on a gas with atomic hydrogen but we also present results for $H_2$ and He gases.

As discussed by Watcharinyanon et al. [19]: *atomic hydrogen exposures on a monolayer graphene grown on the SiC (0001) surface result in hydrogen intercalation. This induces a transformation of the monolayer graphene and a carbon buffer layer to bi-layer graphene without a buffer layer. STM, LEED, and core-level photoelectron spectroscopy measurements have revealed that hydrogen atoms can go underneath the graphene and the carbon buffer layer and bond to Si atoms at the substrate interface. This transforms the buffer layer into a second graphene layer. Hydrogen exposure results initially in the formation of bi-layer graphene islands on the surface. With larger atomic hydrogen exposures, the islands grow in size and merge until the surface is fully covered with bi-layer graphene. An interesting STM study of graphite exposed to atomic hydrogen [20] showed that* **hydrogen intercalated and was stored between the graphite layers and resulted in the formation of graphene blisters**. *Recent STM studies of atomic hydrogen exposures on graphene grown on SiC(0001) [21] and Ir(111) [22] reported that hydrogen adsorbed on the graphene surface without penetrating through the graphene layer into the substrate interface.*

This mechanism of hydrogen intercalation should be the dominating effect driving formation of bi-layer graphene but the effect considered in this work will often be present in real experimental situations and supports the formation of bi-layers as well as influences the distance between bilayers.

We present in Sec. 2 the theoretical model used to calculate the total van der Waals (vdW) potential. The vdW potential can in some limits become attractive and in others repulsive. We also discuss how insertion of different gas particles (hydrogen atoms, hydrogen molecules, and helium atoms) changes the critical concentrations going from attractive to repulsive van der Waals forces. Numerical results are presented in Sec. 3. Here we demonstrate that doping has a weakening effect on the net repulsion. Finally, in Sec.4 we end with summary and conclusions.

**2. Theoretical Section**

To calculate van der Waals forces in multilayer systems we start from the approach explored by Ninham, Parsegian and co-workers [7,23,24] based on the formalism developed by Lifshitz and co-workers [2, 25]. We extend this formalism to incorporate graphene layers at interfaces. We consider a 2D graphene sheet (with polarizability $\alpha_1(q,i\omega)$) adsorbed at the interface between silica (with dielectric function $\varepsilon_1(i\omega)$) and atomic hydrogen gas with dielectric function $\varepsilon_2(i\omega)$. A distance $L$ away from this interface there is a second freestanding 2D graphene sheet (with polarizability $\alpha_2(q,i\omega)$). Outside the second sheet is air ($\varepsilon_3(i\omega)=1$). The geometry is illustrated in Fig. 1. Note that we include spatial dispersion completely in the treatment of the graphene layers, but neglect it in the three regions surrounding the sheets. One of the benefits with 2D layers is that spatial dispersion can be easily incorporated [15] while for 3D layers this is much more complicated [26,27].

We limit the calculation to small separations where we can neglect both retardation and thermal effects. We calculate the non-retarded Casimir (or van der Waals) interaction. In the derivation we solve Maxwell's equations in the there regions and use the standard boundary conditions at the interfaces. The induced carrier densities in the graphene layers are treated as external so the normal components of the **D**-fields are discontinuous at an interface containing a graphene sheet. The discontinuity is $4\pi\rho(\mathbf{q},\omega)$ in cgs units, where $\rho(\mathbf{q},\omega)$ is the 2D Fourier transform of the induced charge density. This density is related to the potential according to $\rho(\mathbf{q},\omega) = \chi(\mathbf{q},\omega) v(\mathbf{q},\omega)$ where $\chi(\mathbf{q},\omega)$ is the electric susceptibility or density-density correlation function, which is related to the polarizability as $\alpha(\mathbf{q},\omega) = -2\pi e^2 \chi(\mathbf{q},\omega)/q$.

At zero temperature we find the energy per unit area is given by,

$$E(d) = \frac{\hbar}{4\pi^2} \int_0^\infty d\omega \int_0^\infty dq\, q \ln\left\{1 - \frac{\left(\frac{\varepsilon_2 - \varepsilon_3}{2} - \alpha_2\right)\left(\frac{\varepsilon_2 - \varepsilon_1}{2} - \alpha_1\right)}{\left(\frac{\varepsilon_2 + \varepsilon_3}{2} + \alpha_2\right)\left(\frac{\varepsilon_2 + \varepsilon_1}{2} + \alpha_1\right)} e^{-2qL}\right\}. \qquad (1)$$

One should observe that our expression for the energy could also be put to use in several other systems. Say for example that $\alpha_2(q,i\omega) = 0$ and $\varepsilon_2(i\omega)$ and $\varepsilon_3(i\omega)$ are taken as the dielectric function of water (or air) and the material in a colloidal tip then one can use this expression to calculate the van der Waals interaction between a graphene coated silica surface in water (or air) and a tip. We will pursue this line of work elsewhere.

We consider the case of two identical undoped graphene sheets with the polarizability given in the literature [15, 28-30],

$$\alpha(q,i\omega) = \left(\frac{-2\pi e^2}{q}\right)\left(\frac{-g}{16\hbar}\right)\frac{q^2}{\sqrt{v^2 q^2 + \omega^2}}. \qquad (2)$$

Here $e$ is the unit electric charge, $v$ is the carrier velocity, which is constant in graphene, and $g$ represents the degeneracy parameter with a value of four (a factor of two for spin and a factor of two for the cone degeneracy). The value of $v$ is *8.73723m/s* [29]. This expression for the polarizability is valid in the undoped case. In the case of doping it becomes much more complicated. We have used the expression derived in Ref. [15], which is expressed in terms of real valued functions of real valued variables.

For the model dielectric function of silica we use that given by Grabbe [31] and for the dielectric function of the atomic gas we use the asymptotic result for low concentrations,

$$\varepsilon_2(i\omega) = 1 + 4\pi n_H \alpha_H(i\omega), \tag{3}$$

where $n_H$ is the concentration of atoms (or molecules) and $\alpha_H(i\omega)$ is the polarizability of atomic hydrogen (or molecular hydrogen or helium atoms) as given by Rauber et al. [32].

## 3. Numerical Results

The important quantity that drives the repulsive van der Waals force is the difference between the dielectric functions of the background materials. Consider first the case when there are no graphene sheets. As has been explored by Parsegian and others the key quantity that determines the sign of the van der Waals potential in triple layer films is the frequency dependent dielectric function (which correspond to the optical properties of the media). We show in Fig. 2 the dielectric function of silica and an atomic hydrogen gas with specified atom concentration. When (as in the figure) $\varepsilon_1 < \varepsilon_2 < \varepsilon_3$ the van der Waals potential turns repulsive [5-7]. It will be repulsive also at low concentrations of atomic hydrogen gas (if one ignores the practical fact that the hydrogen would leak out in the absence of the bi-layer graphene...). However, when the bi-layer graphene is added to the system there will for low hydrogen concentration, and short separations, be an attractive van der Waals potential dominated by the interaction between the graphene sheets. The total van der Waals interaction energy between two undoped graphene sheets (one adsorbed to $SiO_2$, one freestanding a distance $L$ from the surface for two different concentrations ($n_H$) of hydrogen atoms between the sheets) is shown in Fig. 3. Beyond a critical distance the vdW potential turns repulsive with a maximum at a well-defined distance between the sheets.

In Fig. 4 we compare, with same particle concentration, the distance dependence for a gas of atomic hydrogen, molecular hydrogen, and helium atoms. We observe similar trends for all three gases. What makes atomic hydrogen special is the fact that it interacts with the surface in a well-known fashion helping in the formation of bi-layer graphene.

In Fig. 5 we study the effect of doping of the graphene sheets. The geometry is the same as in Fig. 4 with atomic hydrogen between the sheets. The curves are for different doping concentrations. We find that doping reduces the repulsive Van der Waals interaction, but we stress that even in the presence of high doping concentration it is possible to have repulsive Van der Waals interaction.

In Fig. 6 we consider again the total van der Waals interaction energy between two undoped graphene sheets (one adsorbed to $SiO_2$, one freestanding at a fix distance $L$ from the surface) as a function of concentration of hydrogen atoms between the sheets. There is a critical concentration where the total van der Waals potential turns repulsive for a specified $L$. The attraction at high hydrogen concentrations comes about because the dielectric function of the gas at high concentration becomes larger than that of $SiO_2$. It is really not realistic to consider the upper concentration range as a diluted gas so we focus on the lower concentration range where repulsion sometimes occurs.

In the standard treatment of the interaction between graphene sheets one only includes contributions from the graphene valence bands ($\pi$- bonds) [15]. We have done so also in this work. To get an estimate of how good this approximation, neglecting the $\sigma$- bonds, is we have calculated the interaction energy between two thin diamond films and compared the results to the corresponding energy between two graphene sheets, see Fig. (7). The thickness of the films was chosen so that each film contains the same concentration of σ-electrons per unit area as in graphene. In diamond [33], carbon atoms are connected in a fourfold-coordinated tetrahedral bonding structure (sp3 hybrids, four σ- bonds per carbon atom); while in graphene, the atoms are in a threefold-coordinated hexagonal bonding planar structure (sp2 hybrids, three σ– bonds plus one π– bond per carbon atom).

This calculation was performed in the same way as in an earlier work on thin metal films [34]. The dielectric function for diamond was calculated as in the work of Gioti and Logothetidis[35]. As we see in the figure the contribution from the π- bonds is at least a factor of three larger than that from the σ- bonds.

## 4. Summary and conclusions

We have demonstrated that addition of atomic hydrogen, molecular hydrogen, and helium atoms between bi-layer graphene sheets (one adsorbed on a silica surface and the other some nanometer from the surface) can give rise to repulsive van der Waals forces between the sheets.

Graphene sheets are anisotropic which will also influence the van der Waals force between two graphene sheets. Parsegian and co-workers have calculated van der Waals forces as a function of the orientation and dielectric anisotropy [7]. The anisotropic corrections to the dielectric functions can be calculated, as demonstrated by Ninham and Sammut [36] (in a generalization of Lord Rayleigh's earlier extension of the Clausius-Mossotti relation). Smith and Ninham [37] performed a calculation of forces between anisotropic liquid crystals.

A point raised by Smith and Ninham [37] is that van der Waals forces with anisotropic bi-layers can go from attractive to repulsive depending on orientation also in the absence of intervening gas. This is a topic that we intend to explore further. However, the effect studied in this work will in experimental situations be present to larger or smaller degree when atomic hydrogen is pumped inside a bi-layer graphene system adsorbed on a silica surface. We get a similar effect with addition of other gases. It is well known experimentally that not only atomic hydrogen can cause bi-layer graphene. Similar effects occur with lithium intercalation [38]. Hydrogen intercalation provides p-doping [39] while Li intercalation [38] provides strong n-doping which together may provide interesting avenues to nano-scale electronics. We plan in the near future to investigate what happens with van der Waals forces with p-doped and n-doped bi-layer graphene near a SiC surface.

The system we have treated in this work has similarities with that discussed by Munday, Capasso and Parsegian in Ref. [40]. In experiments they found repulsive Casimir- and van der Waals-potentials that can cause quantum levitation of one surface above another in a fluid or gas. This could also lead to the suppression of stiction and to ultra-low friction devices and sensors.

## Acknowledgements

The research was sponsored by the VR-contract No:70529001 which is gratefully acknowledged. We thank Prof. Barry W. Ninham for useful comments.

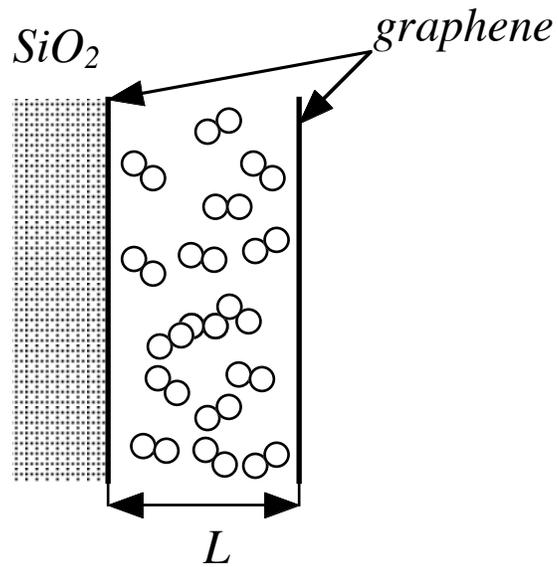

Fig. 1. Geometry considered in this work: Two graphene sheets, undoped or doped, a distance $L$ apart, one adsorbed to a $SiO_2$ substrate and the other freestanding; the space between them is filled with a gas.

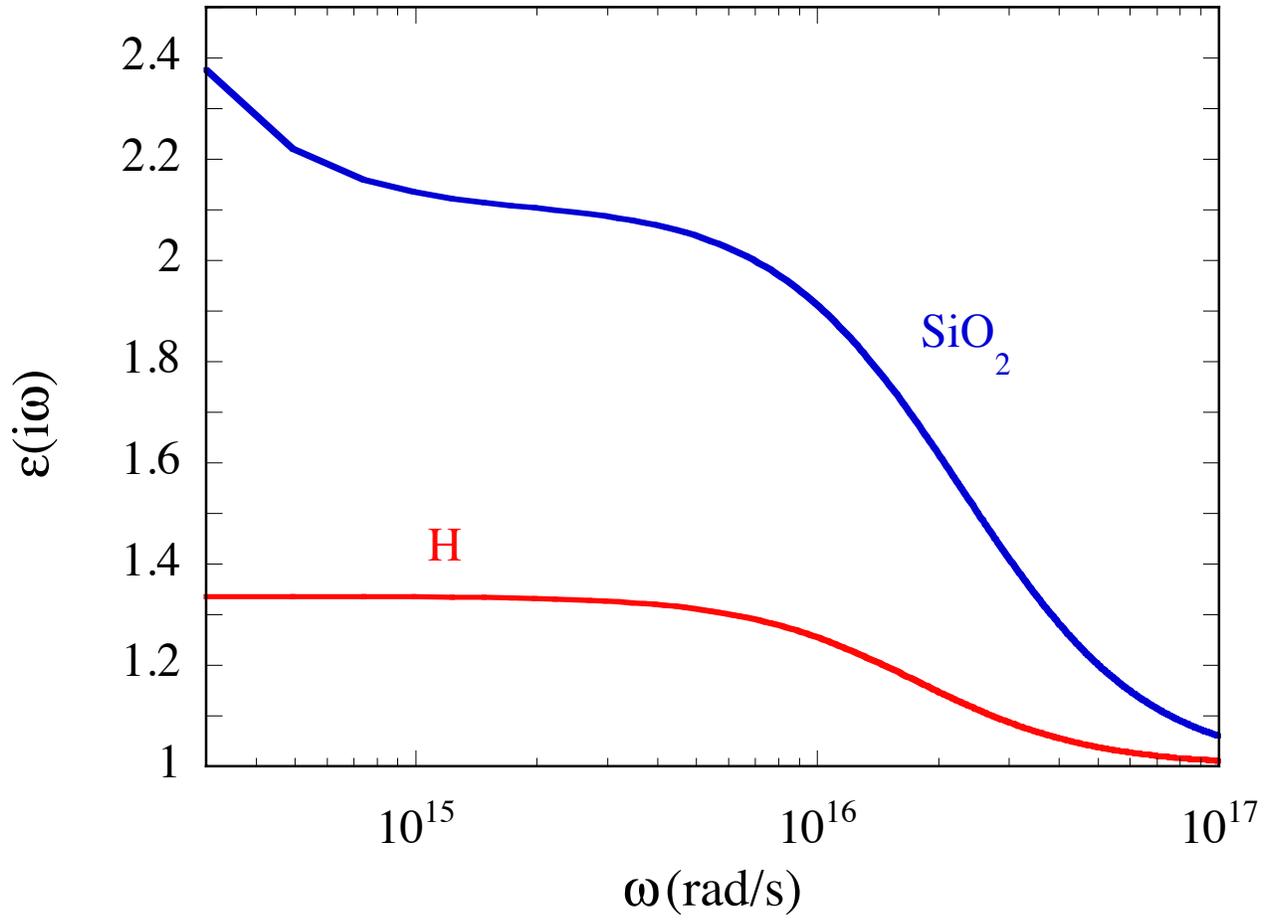

Fig. 2. (Color online) The dielectric function at imaginary frequencies for $SiO_2$ (upper curve) and atomic hydrogen gas at a concentration of $n_H = 4 \times 10^{22}$ cm$^{-3}$ (lower curve).

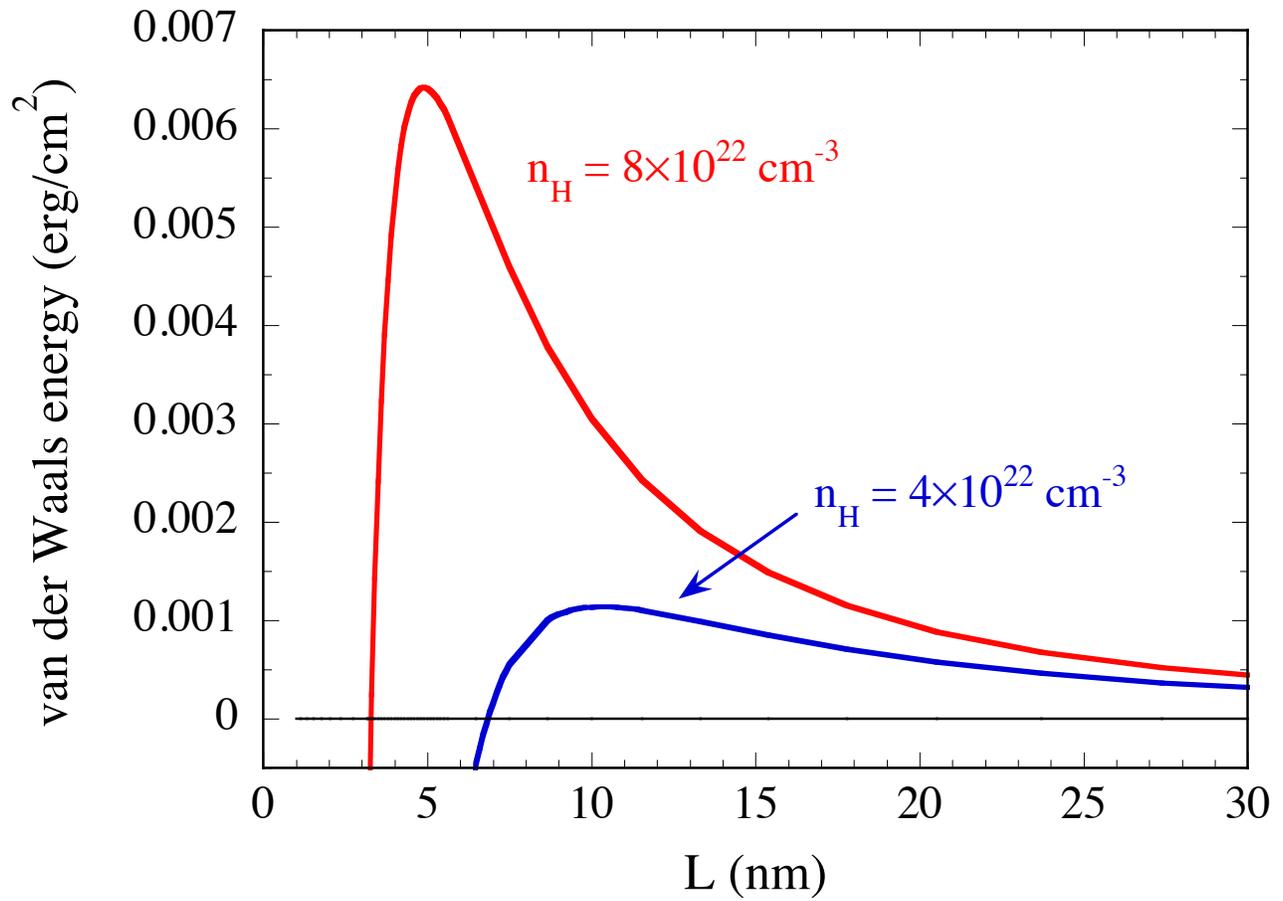

Fig. 3. (Color online) The van der Waals interaction energy between two undoped graphene sheets (one adsorbed to SiO$_2$, one freestanding a distance $L$ from the surface) considering two different concentrations ($n_H$) of hydrogen atoms between the sheets.

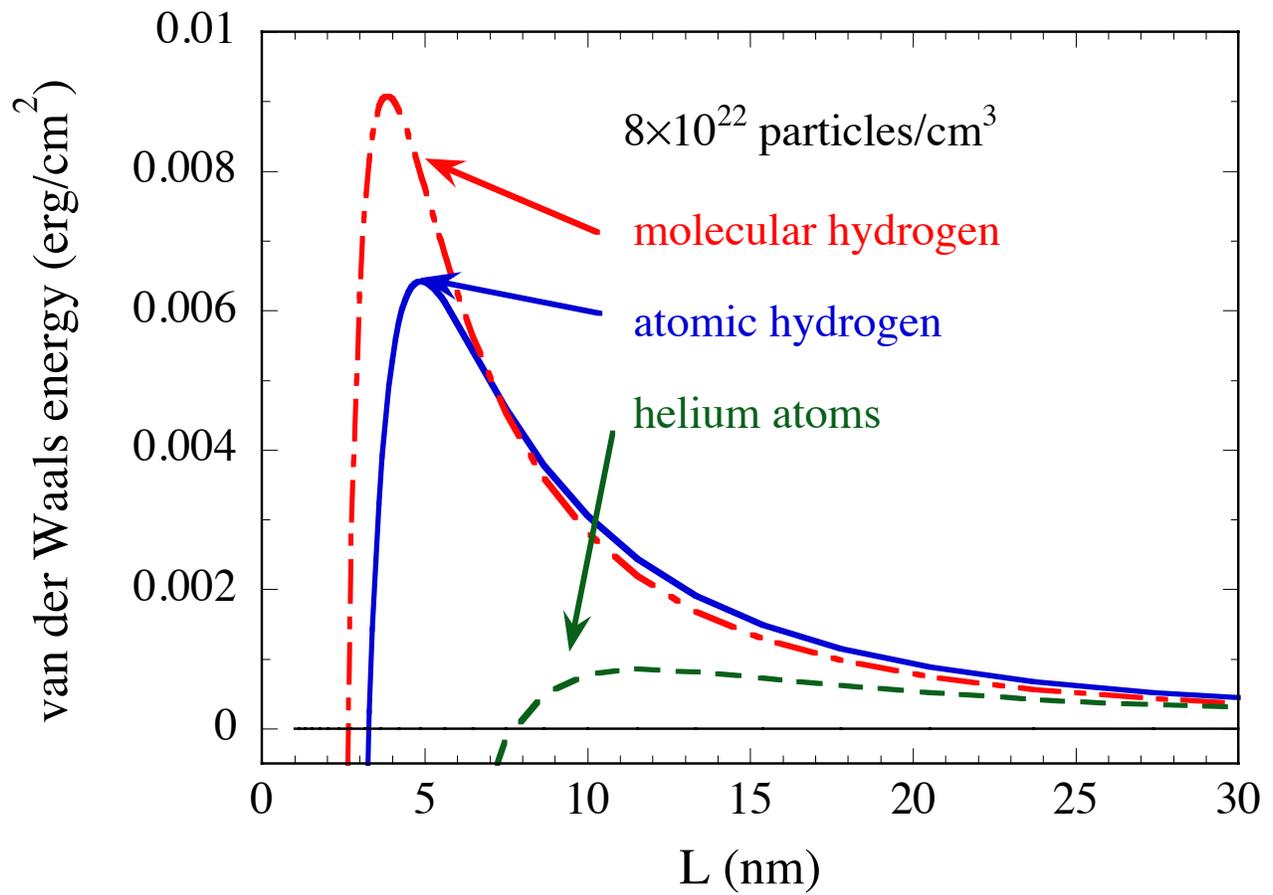

Fig. 4. (Color online) The van der Waals interaction energy between two undoped graphene sheets (one adsorbed to $SiO_2$, one freestanding a distance $L$ from the surface) considering a specific concentration of hydrogen atoms, hydrogen molecules, and helium atoms between the sheets.

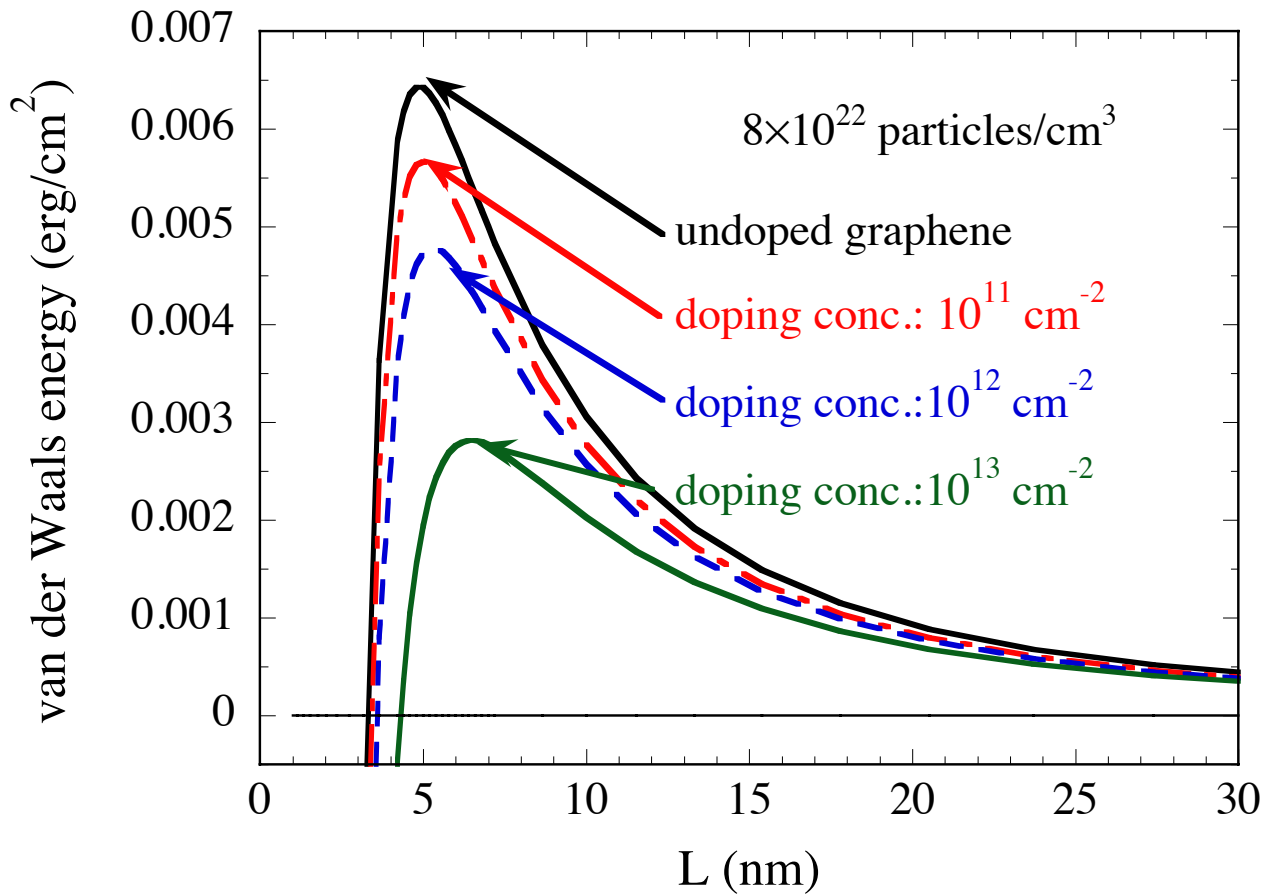

Fig. 5. (Color online) The van der Waals interaction energy between two graphene sheets (one adsorbed to $SiO_2$, one freestanding a distance $L$ from the surface) for a specific concentration of atomic hydrogen between the sheets. The upper curve is for undoped graphene while the lower curves are for different doping concentration. We see that the doping reduces the repulsive Van der Waals interaction.

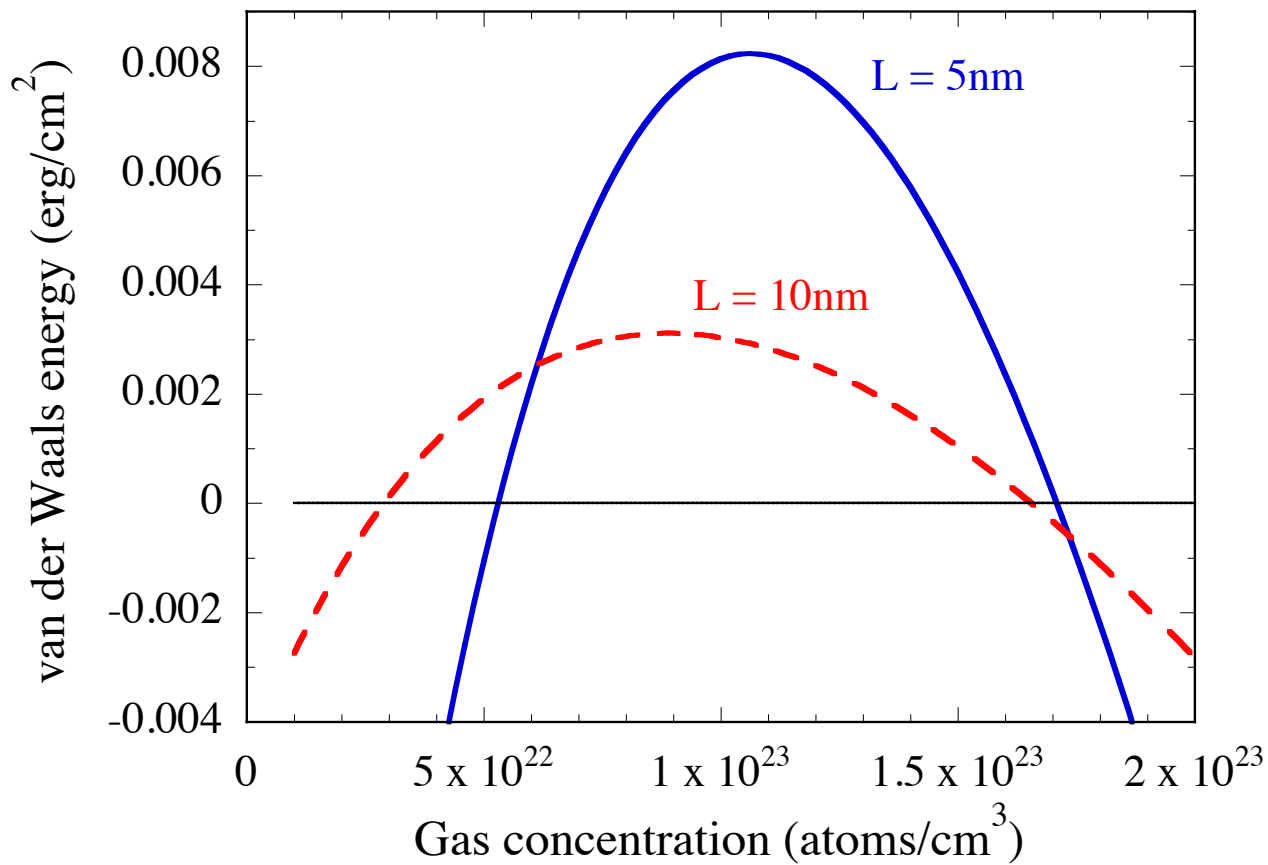

Fig. 6. (Color online) The total van der Waals interaction energy between two undoped graphene sheets (one adsorbed to $SiO_2$, one freestanding a distance $L$ from the surface) as a function of concentration of hydrogen atoms between the sheets.

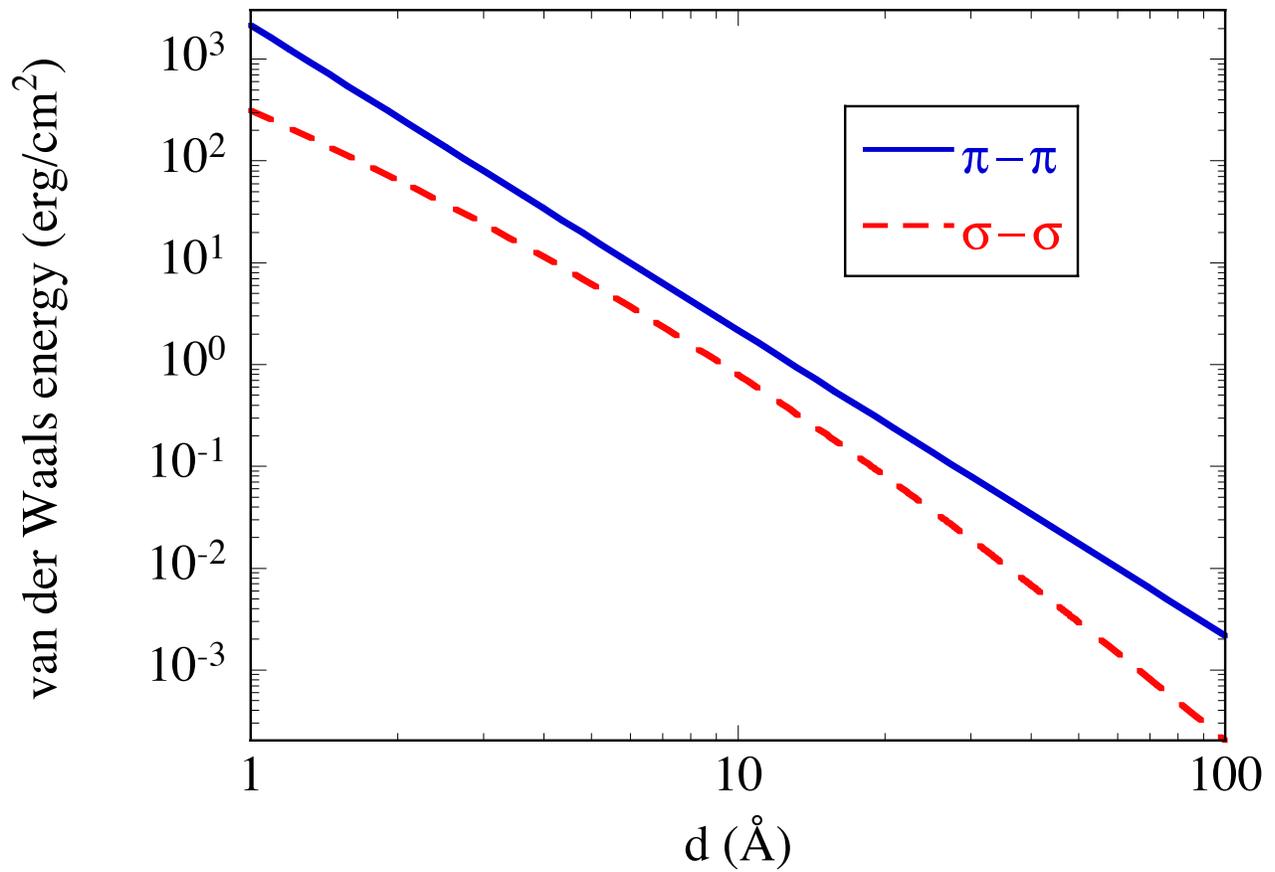

Fig.7 (Color online) Van der Waals energy between two free-standing graphene sheets. The solid curve is the result from the standard approach only including contributions from the valence bands (π- bonds) [15]. The dashed curve is an estimated contribution from the σ- bonds.